\documentclass[a4paper,11pt,custombib]{article}
\usepackage{pos}
\usepackage{lineno}

\title{One model to rule them all: magnetic braking from CVs to low-mass stars}

\author*[a]{Arnab Sarkar}
\author[b,c,d]{Hongwei Ge}
\author[e]{Lev Yungelson}
\author[f]{Christopher A. Tout}

\affiliation[a,f]{Institute of Astronomy, University of Cambridge,\\
The Observatories, Madingley Road, Cambridge CB3 OHA, UK}

\affiliation[b]{Yunnan Observatories, Chinese Academy of Sciences,\\
Kunming 650216, China}
\affiliation[c]{Key Laboratory for Structure and Evolution of Celestial Objects, CAS,\\
	Kunming 650216, China}
\affiliation[d]{International Centre of Supernovae, Yunnan Key Laboratory,\\
	Kunming 650216, China}

\affiliation[e]{Institute of Astronomy of the Russian Academy of Sciences,\\ 48 Pyatnitskaya Str.,119017 Moscow, Russia
}
\emailAdd{as3158@cam.ac.uk}

\abstract{
We present the results of the study of the secular evolution of donor stars in cataclysmic variables (CVs) and AM Canum Venaticorum (AM CVn) stars with our double dynamo (DD) formalism of angular momentum loss (AML) by magnetic braking (MB). In our DD model, MB is driven by an interplay between two $\alpha-\Omega$ dynamos, one in the convective envelope and the other at the boundary of the radiative core and the convective envelope. We show that (1) our MB model reproduces the period gap ($2\lesssim P_\mathrm{orb}/\,\mathrm{hr}\lesssim3$) and the period minimum spike  ($P_\mathrm{orb}\approx 80\, \mathrm{min}$) in the distribution of non-magnetic CVs, (2) evolved CVs, where the donor star commences Roche lobe overflow (RLOF) close to or just beyond the end of the main-sequence, populate the region in and beyond the period gap, and are more likely to be detected at $P_\mathrm{orb}\geq 5.5 \,\mathrm{hr}$. This likely contaminates the mass-radius fit of long-period CV donors. We show that (3) some of the evolved CVs become AM CVn stars with $10\lesssim P_\mathrm{orb}/\,\mathrm{min}\lesssim 65$. Their evolution, driven by $\mathrm{AML_{MB}}$ and AML by gravitational radiation (GR, $\mathrm{AML_{GR}}$), leaves them extremely H-exhausted to the point of being indistinguishable from AM CVn systems formed via the He-star and the White Dwarf (WD) channels in terms of the absence of H in their spectra. We further show that (4) owing to the presence of a significant radiative region, intermediate-mass giants/sub-giants, which are progenitors of AM~CVn stars formed through the He-star channel, may undergo common envelope evolution that does not behave classically, (5) several AM CVn systems with extremely bloated donors, such as Gaia14aae, ZTFJ1637+49 and SRGeJ045359.9+622444 do not match any modelled trajectories if these systems are modelled only with $\mathrm{AML_{GR}}$ without incorporating $\mathrm{AML_{MB}}$, (6) the uncertainties in MB estimates greatly affect modelling results. This, in turn, affects our efforts to distinguish between different AM CVn formation channels and their relative importance. Finally, we find that (7) a similar MB prescription also explains the spin-down of single, low-mass stars. 

}
\FullConference{%
  The Golden Age of Cataclysmic Variables and Related Objects VI\\
  4-9 September, 2023\\
  Palermo, Italy}

\tableofcontents

\begin{document}
\maketitle

\section{Introduction}
Cataclysmic variables are a class of interacting binary systems consisting of a mass-transferring secondary star along with a mass-accreting white dwarf (WD) primary \citep[][]{2003cvs..book.....W}. The secular evolution of CVs is driven by the loss of angular momentum from the system, which leads to the donor filling its Roche lobe and commencing mass transfer. According to the canonical model of CV evolution, for longer orbital periods ($P_\mathrm{orb}\gtrsim 3\,\mathrm{hr}$), the primary mode of angular momentum loss is some sort of magnetic braking (MB) owing to a stellar wind from the donor star. A dearth of observed mass transferring CVs between $2\,\lesssim  P_\mathrm{orb}/\mathrm{hr}\lesssim3$ (called the period gap) led to the interrupted magnetic braking paradigm \citep[][]{1983A&A...124..267S} wherein MB stops abruptly when the donor becomes fully convective (at $P_\mathrm{orb}\approx 3\,\mathrm{hr}$).  At this point, the donor, which had been driven out of thermal equilibrium because of mass loss, contracts, causing the cessation of mass transfer. From here on only gravitational radiation remains as a mechanism for angular momentum loss. Mass transfer begins again only when the Roche lobe catches up with the convective donor at $P_\mathrm{orb}\approx2\,\mathrm{hr}$. The evolution of CVs is also governed by the interplay between the donor's mass-loss timescale $\tau_\mathrm{ML} \approx M_2/\Dot{M}_2$ and its Kelvin-Helmholtz timescale $\tau_\mathrm{KH}\approx GM_2^2/R_\ast L_\ast$, where $M_2$, $R_\ast$ and $L_\ast$ are the donor's mass, radius and luminosity. As long as $\tau_\mathrm{ML} \gg \tau_\mathrm{KH}$, the donor is able to maintain thermal equilibrium and behave like a standard main-sequence star. However, when $\tau_\mathrm{ML} \approx \tau_\mathrm{KH}$ mass transfer leads to an increase in the donor's size and $P_\mathrm{orb}$ increases in response to it. This leads to a period minimum $P_\mathrm{min}$ as the donor transforms from a shrinking MS star to an expanding, partially degenerate one \citep[][]{1981ApJ...248L..27P, 1982ApJ...254..616R}. 

AM Canum Venaticorum (AM CVn) stars are a class of semi-detached binaries with extremely short orbital periods, $10\lesssim~P_\mathrm{orb}/\,\mathrm{min}\lesssim 65$. Although closely related to CVs, these systems have shorter orbital periods and usually lack H in their spectrum (see \citealt{Solheim2010} for a detailed review). They are usually modelled as an evolved star transferring mass to a WD accretor. Owing to their short orbital periods, these systems are strong gravitational wave sources \citep{Kupfer2016}. Three possible formation channels for AM CVn systems have been \textcolor{black}{proposed} (\citealt{2014LRR....17....3P}). These differ from each other based on the number of common envelope evolution (CEE) phases the primordial main-sequence (MS) binary goes through and the nature of the donor star.   In the first formation channel, known as the WD channel, the donor is a He WD which commences Roche lobe overflow (RLOF) and transfers mass to a more massive \textcolor{black}{carbon-oxygen} (C/O) WD after going through two common envelope (CE) phases (\citealt{1967AcA....17..287P}). The second channel is known as the He-star channel, wherein the donor commences RLOF as either a non-degenerate or semi-degenerate He-rich or He-burning star and transfers mass to a WD after going through two CE phases (\citealt{1986A&A...155...51S}). The final channel is known as the Evolved CV channel \textcolor{black}{in which} an evolved MS star commences stable RLOF after going through a single CE phase and transfers mass to a WD accretor \textcolor{black}{while} in the Hertzprung gap (between the end of its MS and the beginning of its \textcolor{black}{ascent of the} red giant branch,  \citealt{1985SvAL...11...52T, 1987SvAL...13..328T}). 

We review our recent developments in modelling the secular evolution of CVs and, in particular, AM~CVn stars using the Double Dynamo (DD) model of MB first proposed by \cite{Zangrilli1997}, and our efforts to extend this model to address the spin-down of single, low-mass stars. In Section~\ref{sec:DD}, we describe the DD model and how it operates in CVs. In Section~\ref{sec:evolvedcv}, we discuss how evolved CVs populate long orbital periods in the CV distribution. In Section~\ref{sec:2step_he_star} we show how He-star progenitors may not undergo classical (dynamical time-scale) common envelope evolution. In Section~\ref{sec:MB_imp} we illustrate the importance of incorporating MB into the modelling of AM CVn stars as well as the effect of MB uncertainties in theoretical results. In Section~\ref{sec:M-dwarfs} we highlight our efforts to use our MB formalism to explain the spin-down of fully convective M-dwarfs. We summarise our results and conclude in Section~\ref{sec:conclusion}.

\section{The double-dynamo model of magnetic braking in cataclysmic variables}
\label{sec:DD}
\citet{ST}, building on the model of \cite{Zangrilli1997}, reproduced the period gap and the period minimum in CV distribution using a MB mechanism with two $\alpha-\Omega$ dynamos operating in the donor ($M_\ast\lesssim 1.4M_\odot$), one at the boundary 
of the radiative core and the convective envelope and the other in the convective envelope. The working of the two dynamos and the equations governing the MB have been explained in detail by \cite{ST}. Here we only highlight the orbital evolution of the CV owing to angular momentum loss (AML) by MB and gravitational radiation (GR).

\begin{figure*}
\centering
\includegraphics[width=0.7\textwidth]{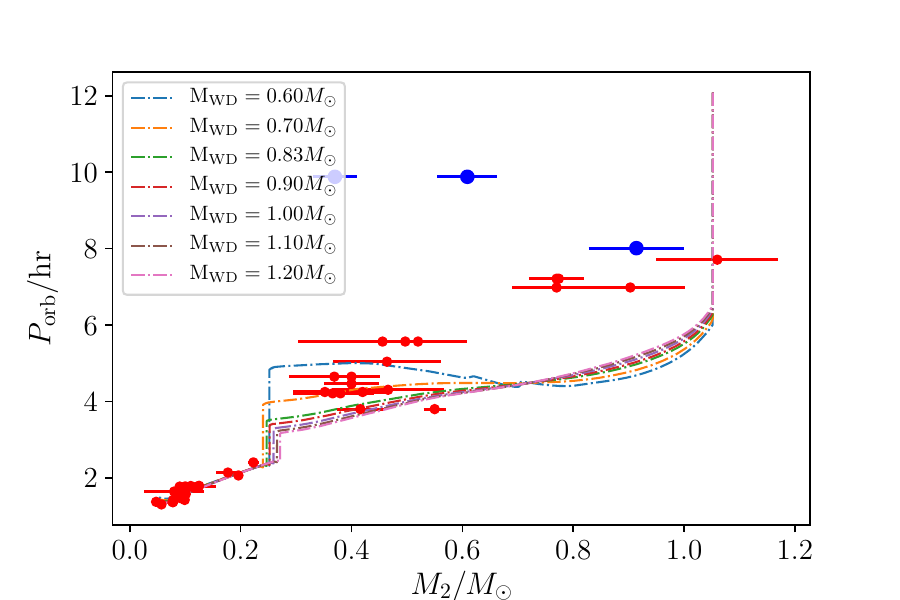}
\caption{Full evolutionary tracks for $M_{2,0}=1M_\odot$ for different WD masses $M_1$ plotted with observed CV data collected by \protect\cite{Ge2015}. The points in red are CVs with non-magnetic accretors. The points in blue are intermediate polars which are not modelled in this work. This figure is taken from \cite{ST}. }
\label{fig:ST_PM}
\end{figure*}

The orbital evolution of CVs has been illustrated in Fig.~\ref{fig:ST_PM}. We begin with a detached system with a donor of mass $M_{2,0}=1M_\odot$ and accretors $M_2=M_\mathrm{WD}$ of different masses with an initial orbital period $P_\mathrm{orb}=12\,$hr.  Initially, the orbit shrinks as a response to AML till $P_\mathrm{orb}\approx 6\,$hr when the donor star fills its Roche lobe and commences mass-transfer to the WD accretor. We assume that the mass transfer is fully non-conservative and that the accreted mass onto the WD is lost in nova eruptions with the specific angular momentum of the WD. Till $M_2\approx0.25M_\odot$ (depending on $M_\mathrm{WD}$, and consequently $\Dot{M}_\mathrm{RLOF}$), AML is dominated by MB due to the boundary layer dynamo. However, this AML stops abruptly because the donor star becomes fully convective. At this point, the donor, which had been bloated by the strong mass loss, shrinks back into its Roche lobe and the system detaches. The orbit is now shrunk by $\mathrm{AML}_\mathrm{GR}$ and $\mathrm{AML}$ due to the convective dynamo. Mass transfer resumes at $P_\mathrm{orb}\approx 2\,$hr. The strength of MB owing to the convective dynamo increases with decreasing mass (equation~9 of \citealt{ST}) and $\tau_\mathrm{KH}\approx\tau_\mathrm{ML}$ at which the system attains its minimum  $P_\mathrm{orb}\approx 82\,$ min.

We have also estimated the probability of {having} a mass transferring system with a given $P_\mathrm{orb}$ by creating a probability distribution histogram. We divide the orbital period space $P$ evenly in the range $P\in[0,6]\,\mathrm{hr}$ and define the probability $\xi$ of a system being found within a given bin $P\leq P_\mathrm{orb}< P+\mathrm{d}P$ as

\begin{figure}
\centering
\includegraphics[width=0.7\textwidth]{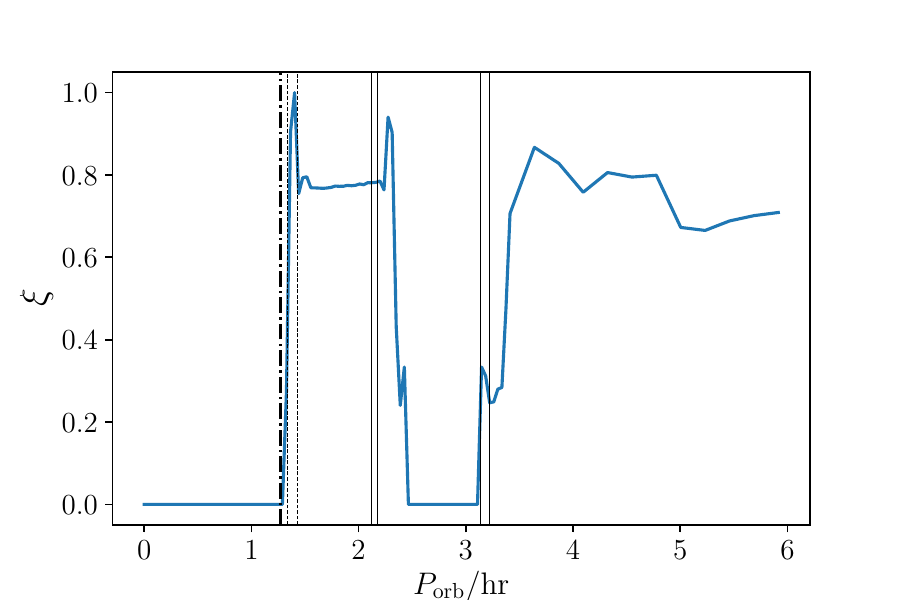}
\caption{The relative probability $\xi$ of a mass transferring system {existing} with a given $P_\mathrm{orb}$ scaled as per the WD distribution function from fig.~6 of \protect\cite{Wijnen2015}. The thin solid black lines are the lower and upper ends of the period gap given by $P_\mathrm{orb,pg,lower}\:=\:2.15\pm0.03\,\mathrm{hr}$ and $P_\mathrm{orb, pg,upper}\:=\:3.18\pm0.04\,\mathrm{hr}$ which we adopt from \protect\cite{Knigge2006}. The dotted black line is the period minimum spike $P_\mathrm{orb, min}\:\in\:[80,86]\,\mathrm{min}$ reported by \protect\cite{Gnsicke2009}. The thick black dash-dotted line is the minimum period $P_\mathrm{orb, min}\:=\:76.2\pm0.03\;\mathrm{min}$ reported by \protect\cite{Knigge2006}. This figure is taken from \cite{ST}. }
\label{fig:pt_scaled}
\end{figure}

\begin{center}
\begin{equation}
\label{prob}
\xi \propto {t_\mathrm{max} - t_\mathrm{min}}
\end{equation}
\end{center}
where $t_\mathrm{min}$ is the time when the system enters the $P$ bin and $t_\mathrm{max}$ is the time when the system leaves it. So that $\xi$ for a given $P$ bin is higher if the system stays in that bin for longer. Because the evolution of $P_\mathrm{orb}$ depends on the WD mass (Fig.~\ref{fig:ST_PM}), we generate a probability distribution histogram by scaling each WD-dependent trajectory with the distribution of WDs in CVs \citep[see fig.~6 of][who use the observed sample of \citealt{2011A&A...536A..42Z}]{Wijnen2015}. This is shown in Fig.~\ref{fig:pt_scaled}. It can be seen that we reproduce the orbital period distribution of CVs of \cite{Knigge2006} and \citet[see their figs~4 and 2 respectively]{Gnsicke2009} taken from the catalogue of \cite{2003A&A...404..301R} and that the period gap and the period minimum spike discussed by \cite{Gnsicke2009} are reproduced quite well.

\section{The systems in and beyond the period gap are evolved CVs}
\label{sec:evolvedcv}

\begin{figure*}
\centering
\includegraphics[width=0.7\textwidth]{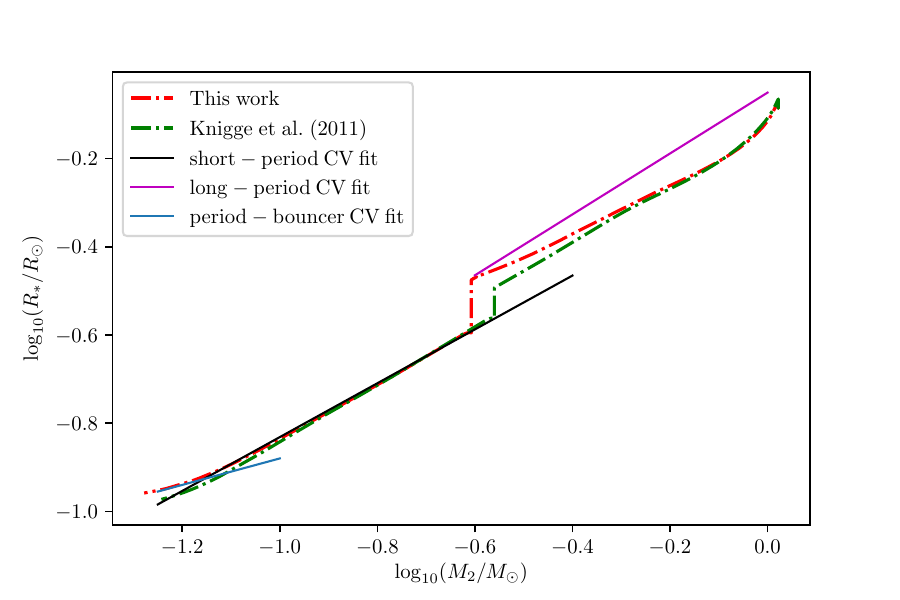}
\caption{$M$ to $R$ relationship of the donor star for our DD model and the model described by \protect\cite{Knigge2011} for a system with $M_2=1M_\odot$ and $M_1=0.83M_\odot$. The short-period CV fit (solid black line), the period-bouncer CV fit (blue line) and the long-period CV fit (magenta line) are the $M$ to $R$ best fits by \protect\cite{Knigge2011}. This figure is taken from \cite{ST}.}
\label{fig:MR}
\end{figure*}

\begin{figure*}
\centering
\includegraphics[width=0.7\textwidth]{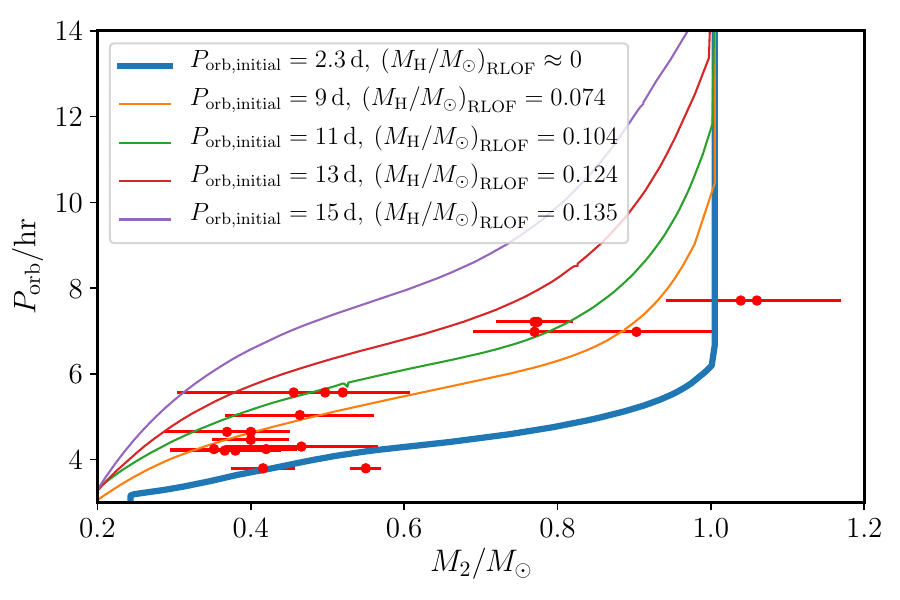}
\caption{Evolutionary tracks of a $M_2=1M_\odot$ and $M_1=1M_\odot$ system in the $(M_2,P_\mathrm{orb})$ plane for different initial orbital periods $P_\mathrm{orb,initial}$ plotted with observed CV data collected by \protect\cite{Ge2015}. \textcolor{black}{The canonical CV is the thick blue line.} \textcolor{black}{Among the evolved systems, larger $P_\mathrm{orb,initial}$ leads to the commencement of RLOF later, leading to more shell H-burning in the subgiant phase of the donor, yielding a larger H-exhausted core. The mass of the H-exhausted core at the beginning of RLOF is also mentioned for each system.} This figure is taken from \cite{SGT_AM}.}
\label{fig:PM_SGT}
\end{figure*}

Fig.~\ref{fig:ST_PM} shows that our trajectories do not agree well with CVs observed with $P_\mathrm{orb}\gtrsim 5\,$hr. This is also evident in the donor mass-radius relationship comparison in Fig.~\ref{fig:MR}. While there is agreement between our model and the empirical fit of \cite{Knigge2011} for short-period CVs and period bouncers (CVs beyond their period minimum), the empirical fit for long-period CVs predicts larger donor radii for a given donor mass, in disagreement with our MB model as well as that of \cite{Knigge2011}. We suspect that these are evolved CVs, such that their donors commence RLOF close to or beyond the terminal-age main sequence. To test this, we plot the evolution trajectories of CVs where the donor has a small H-exhausted core when RLOF begins. This is illustrated in Fig.~\ref{fig:PM_SGT}. It is seen that CVs with evolved donors commence mass transfer at larger periods owing to their radii being bigger than an unevolved CV (canonical CV shown as a thick blue line). The donors are more bloated throughout, resulting in a larger $P_\mathrm{orb}$ at a given $M_2$. This effect is larger for more evolved donors. These trajectories agree well with the observations of CVs with $P_\mathrm{orb}\gtrsim5\,$hr. 

\begin{figure*}
\centering
\includegraphics[width=0.9\textwidth]{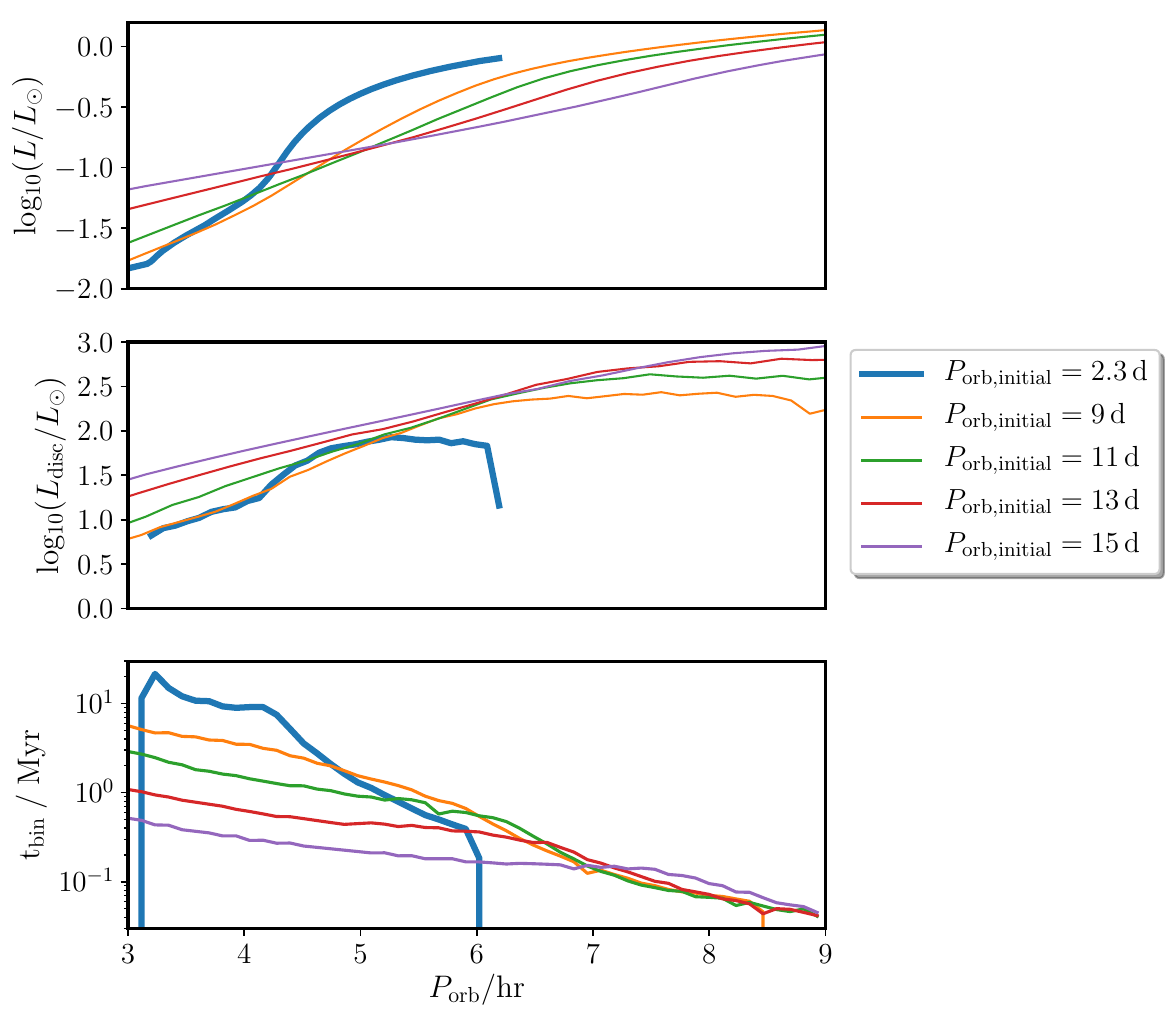}
\caption{\textit{Top and middle}: The evolution of donor luminosity $L_2$ and the disc luminosity $L_\mathrm{disc}$ with $P_\mathrm{orb}$ for the same set of systems as in Fig.~\ref{fig:PM_SGT}. \textit{Bottom}: The time $t_\mathrm{bin}$ spent by each system in a $P_\mathrm{orb}$ bin for the same set of systems as in Fig.~\ref{fig:PM_SGT}. \textcolor{black}{The bin size $\mathrm{d}P_\mathrm{orb}\,=\,0.12\,\mathrm{hr}$.} This figure is taken from \cite{SGT_AM}.}
\label{fig:PT_SGT}
\end{figure*}

We also assess the likelihood of the detection of these systems compared to a canonical CV. In the top two panels of Fig.~\ref{fig:PT_SGT} \textcolor{black}{we show how} the luminosities of the donor $L_2$ and the accretion disc $L_\mathrm{disc}$ evolve with $P_\mathrm{orb}$ after RLOF as a probe of the detection probability of the system. We estimate $L_\mathrm{disc}$ to be
\begin{center}
\begin{equation}
\label{eq:ldisc}
L_\mathrm{disc} \approx \frac{GM_1\lvert\Dot{M}_2\rvert}{R_1},
\end{equation}
\end{center}
where $M_1,\:R_1,\:\mathrm{and}\:\Dot{M}_2$ are respectively the mass, radius and mass accretion rate of the WD primary. \textcolor{black}{For a fixed WD primary, $L_\mathrm{disc}$ only depends on $\Dot{M}_2$. Thus, higher accretion rates will lead to a more luminous disc.} We use $R_1\approx 0.008\,R_\odot$ for $M_1\approx 1\,M_\odot$  \citep{Romero2019}. It can be seen that $L_\mathrm{disc}$ is likely to dominate the luminosity of the system, wherein evolved systems dominate over the canonical CV at almost all orbital periods, illustrating that evolved CVs are more likely to be detected at larger orbital periods. In addition, for $P_\mathrm{orb}\gtrsim 6.5\,\mathrm{hr}$, RLOF has not commenced yet for the canonical CV and so for a $1M_\odot$ donor progenitor we expect mass-transferring systems with $P_\mathrm{orb}\gtrsim 6.5\,\mathrm{hr}$ to just be evolved CVs. The bottom panel of Fig.~\ref{fig:PT_SGT} shows the amount of time each system spends in the bin $P_\mathrm{orb}+\mathrm{d}P_\mathrm{orb}$ (\textcolor{black}{with $\mathrm{d}P_\mathrm{orb}\,=\,0.12\,\mathrm{hr}$}) as another probe of the detection likelihood of the system. The canonical CV spends much more time at $4\,\lesssim P_\mathrm{orb}/\,\mathrm{hr}\lesssim 4.75$ than evolved CVs. However, evolved CVs spend more time at longer $P_\mathrm{orb}$. 
Thus, we can conclude that evolved CVs dominate the orbital period distribution at $P_\mathrm{orb}\gtrsim5.5\,\mathrm{hr}$, as \textcolor{black}{also found by} \cite{Goliasch2015}. As their trajectories closely resemble those of canonical CVs, it is very difficult to distinguish a canonical CV from an evolved CV observationally. We claim that evolved CVs have contaminated the long-period donor mass-radius fit of \cite{Knigge2011}. 
Fig.~\ref{fig:PT_SGT} also shows that evolved CVs do not pass through the period gap as detached systems (\citealt{1985SvAL...11...52T, 1987SvAL...13..328T}) and so are likely to populate the period gap along with canonical CVs born in the gap. These evolved CVs become AM CVn stars through the Evolved CV formation channel at orbital periods less than about 70 min. Their evolution is discussed in Section~\ref{sec:MB_imp}.
 
\section{He-star AM CVn progenitors: the common envelope outcome of subgiants and early red giants may not behave classically}
\label{sec:2step_he_star}

\begin{figure*}
\centering
\includegraphics[width=0.9\textwidth]{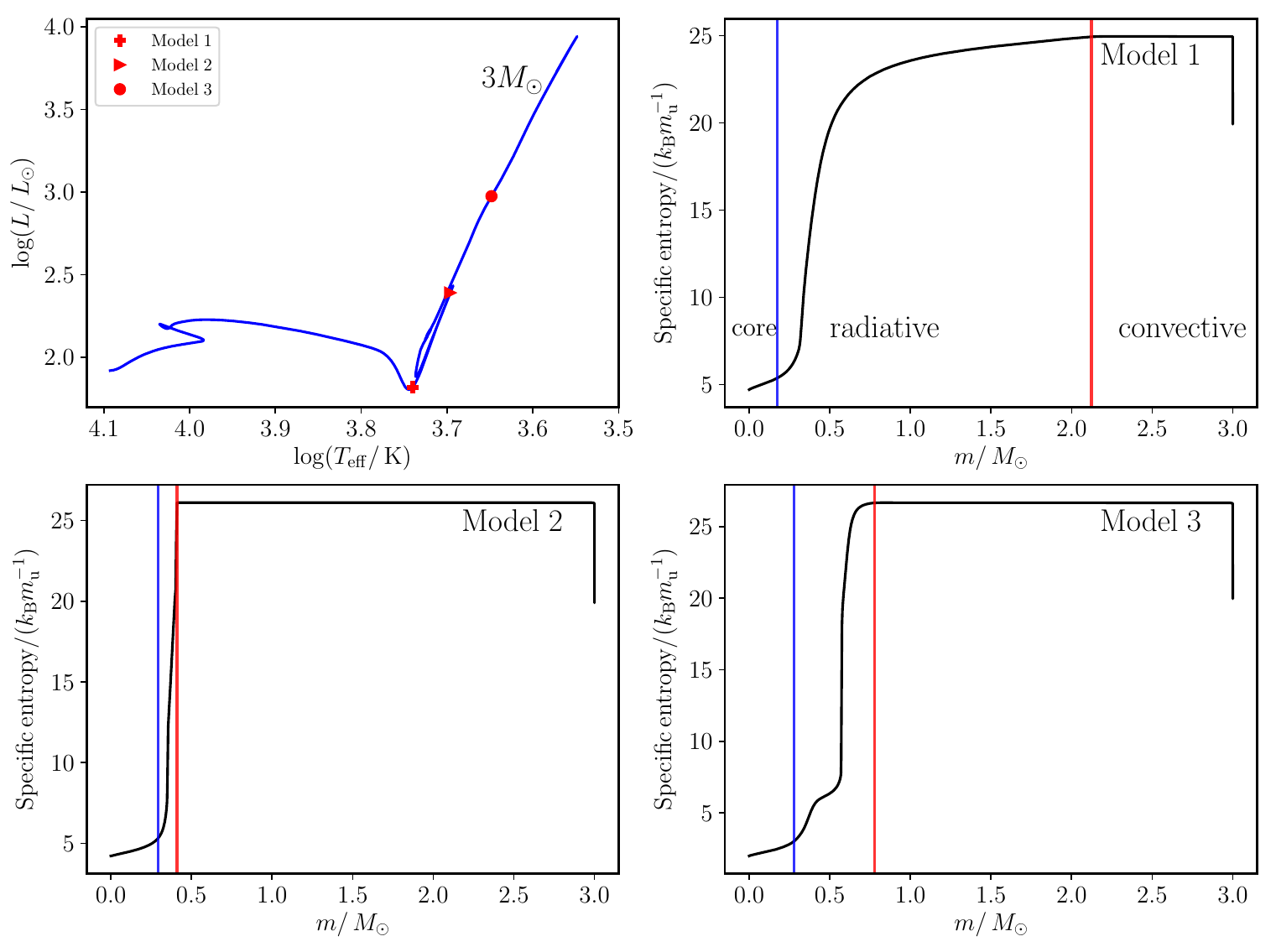}
\caption{The interior structure of a $3M_\odot$ star at different stages of its giant phase. The blue vertical line shows the boundary between the degenerate core and the non-degenerate radiative region and the red
vertical line shows the boundary between the radiative and the convective regions.}
\label{fig:2step_CE}
\end{figure*}

We also explore the possibility of forming binaries consisting of a WD accretor with a semi-degenerate, H-exhausted ($X\leq 10^{-5}$), He-rich ($Y\approx0.98$) donor that are progenitors to the He-star formation channel of AM CVn stars. Because He-star AM CVn systems undergo two common envelope (CE) phases, it is crucial to model the second CE phase that forms the He-rich donor and defines the initial $P_\mathrm{orb}$ of the AM CVn progenitor. So we calculate the outcome of this second CE phase using the energy balance formalism of CE \citep{Han1995} in section~3.1 of \cite{SGT_He}. In this work, we aim to assess whether AM CVn systems discovered by \cite{2022MNRAS.512.5440V} could be formed via the He-star channel. Because these systems show no detection of carbon, we argue that the CE that formed the donor star must have commenced before the ignition of He. The evolutionary stages of the CE progenitor that give us favourable AM CVn progenitors are shown in figs~1 to 4 of \cite{SGT_He}. We show that viable post-CE candidates emerge successfully if CE commences at the subgiant or early red giant phase. However, the energy formalism of CE ejection, which governs the final separation ($P_\mathrm{orb}$) of our AM CVn progenitors predicts a common envelope ejection efficiency $\alpha_\mathrm{CE}>1$. This implies that the orbital energy of the binary system is not sufficient enough to unbind the envelope of the He-star progenitor. It has been proposed recently that the CE phase may not proceed classically (on a dynamical time-scale) in stars that have a substantial radiative region in between their convective envelope and the degenerate core \citep{Hirai2022}. Instead, the CE outcome is a two-step process, the combination of a dynamical time-scale event (that proceeds classically with $\alpha_\mathrm{CE}<1$) plus a thermal time-scale (dynamically stable) event. Although \cite{Hirai2022} describe this for red supergiants, the formalism is quite general and can be implemented in any star that has a substantial radiative region in between its core and the envelope This is the case for the progenitors of our He-star in figs~1 to 4 of \cite{SGT_He}. It is shown for three evolutionary stages of a $3M_\odot$ star in Fig.~\ref{fig:2step_CE}. Modelling the two-step CE with the classical CE formalism is equivalent to setting $\alpha_\mathrm{CE}\gtrsim10$ and so we use this approach to deduce the post-CE He-star mass and orbital separation  \citep{SGT_He}. A detailed and rigorous study will shed more light on the feasibility of the two-step formalism on the CE evolution in subgiants/early red giants.



\section{The importance of magnetic braking in AM CVn evolution and the dependence on magnetic braking uncertainties}
\label{sec:MB_imp}

\begin{figure*}
\centering
\includegraphics[width=0.95\textwidth]{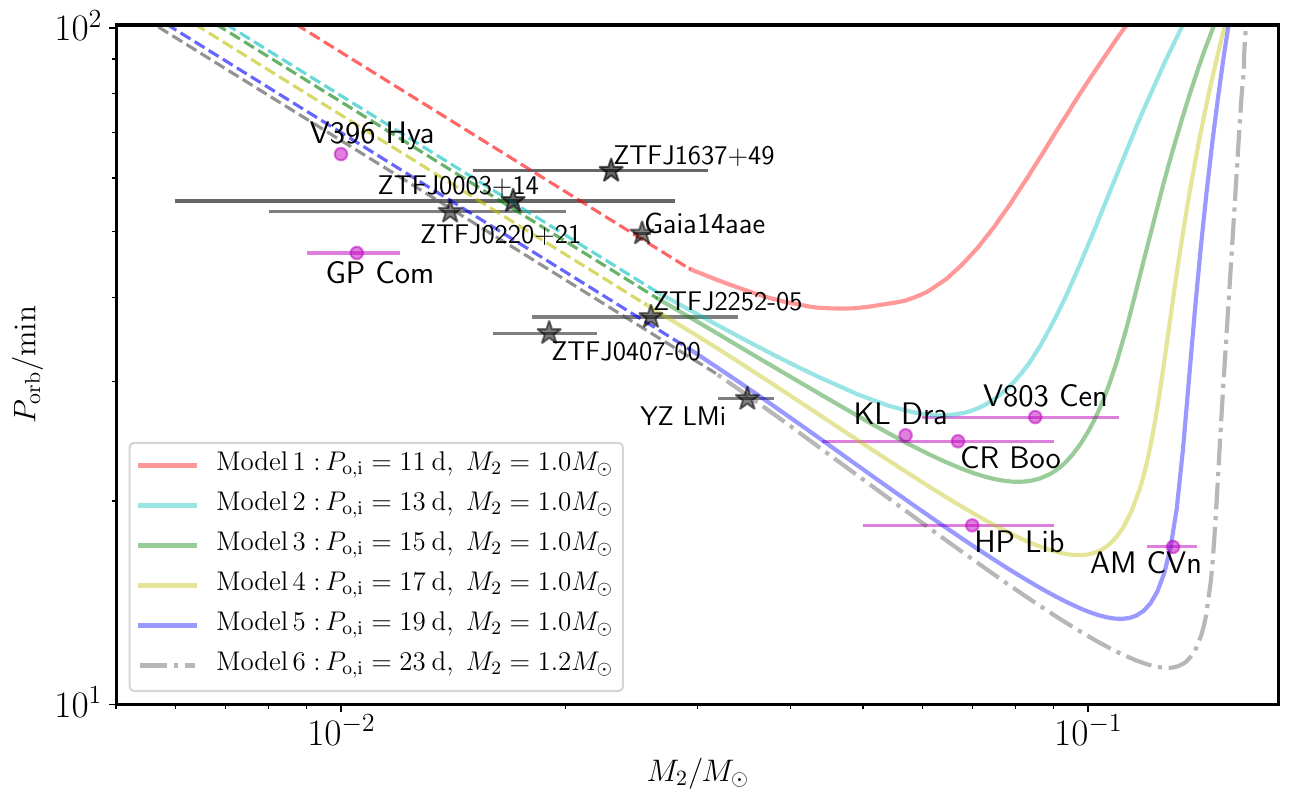}
\caption{The relation of the donor's orbital period and mass in the $(M_2,P_\mathrm{orb})$ plane, for systems with $M_1=1M_\odot$ and different $P_\mathrm{orb,initial}$ ($P_\mathrm{o,i}$). Increasing $P_\mathrm{orb,initial}$ corresponds to more H-exhausted AM CVn stars. The systems with $P_\mathrm{orb,initial}=19\,\mathrm{d}$ and $P_\mathrm{orb,initial}=23\,\mathrm{d}$ commence RLOF at $t\approx t_\mathrm{BGB}$ (the base of the giant branch) for different donor masses. The dashed section in each trajectory is a power law fit of the form $P_\mathrm{orb} \propto (M_2/M_{2,P_\mathrm{orb,min}})^\beta$, where $\beta=-0.6860$. The fit is described in detail in \protect\cite{SGT_AM}. The points in magenta are systems from \protect\cite{Solheim2010}, while the stars in black are systems described by \protect\cite{2022MNRAS.512.5440V}, \protect\cite{Green2018} and \protect\cite{Copperwheat2010}. This figure is taken from \cite{SGT_AM}.}
\label{fig:pm-solvan}
\end{figure*}

\begin{figure*}
\centering
\includegraphics[width=0.95\textwidth]{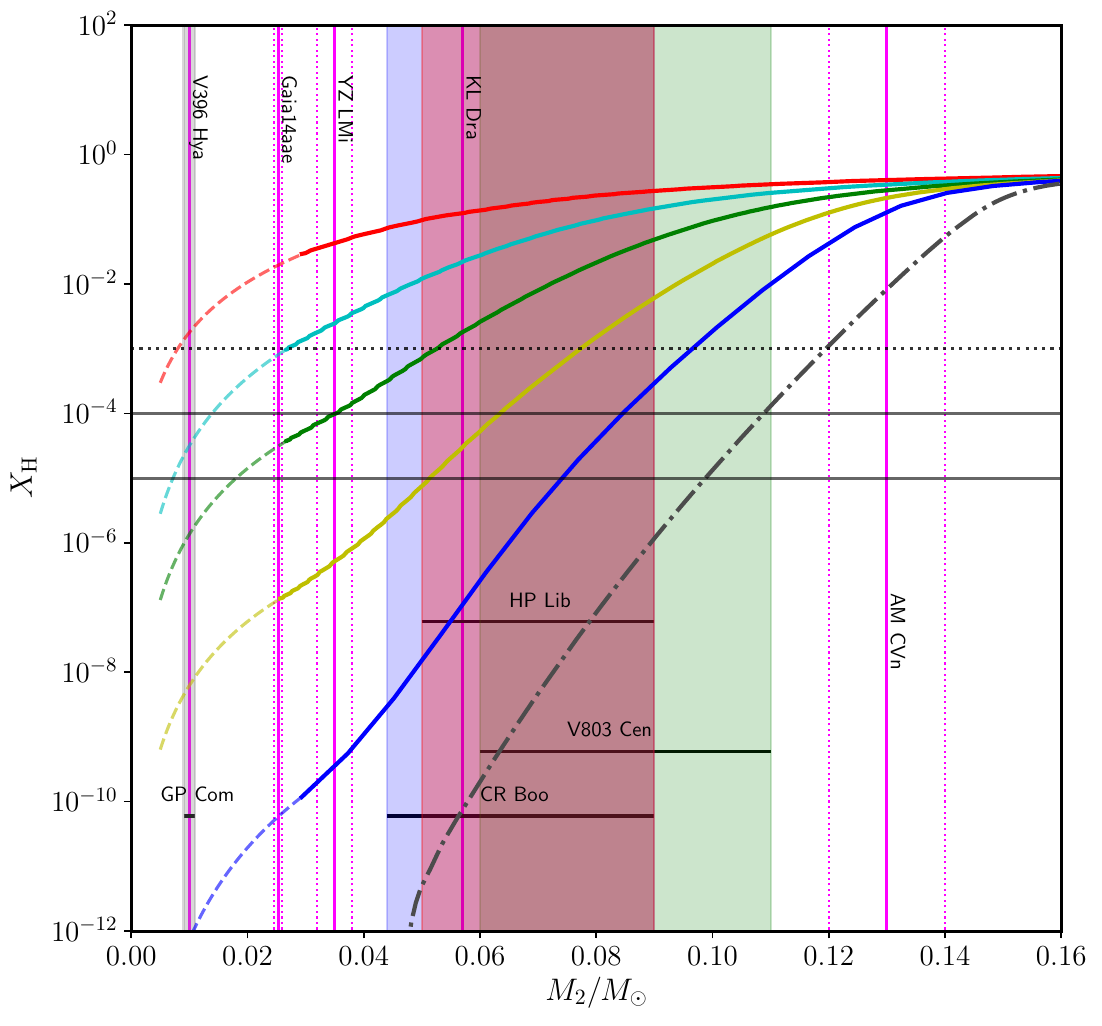}
\caption{The surface H abundance $X_\mathrm{H}$ of the donor with $M_2$ for the same modelled systems as in Fig.~\ref{fig:pm-solvan}. The vertical solid lines and the associated dotted lines in magenta are the donor masses and the error bars of observed systems obtained by \protect\cite{Solheim2010} and \protect\citet[and the references therein]{2022MNRAS.512.5440V}. The shaded regions are donors that have an inferred mass range. V083 Cen (green) has $M_2\in[0.06,\,0.11]M_\odot$, HP Lib (red) has $M_2\in[0.05,\,0.09]M_\odot$, CR Boo (blue) has $M_2\in[0.044,\,0.09]M_\odot$, and GP Com (grey) has $M_2\in[0.009,\,0.012]M_\odot$. Horizontal lines in black denote $X_\mathrm{H}=10^{-4}$ and $10^{-5}$ while the dotted one denotes $X_\mathrm{H}=10^{-3}$. $X_\mathrm{H}\leq10^{-4}$ is the requirement of a system to be H-exhausted \citep{Green2019}. This figure is taken from \cite{SGT_AM}.}
\label{fig:x}
\end{figure*}

In this section, we present our results of modelling evolved CVs beyond their period minimum and short-period He-star plus WD binaries when they become AM CVn stars formed through the Evolved CV channel \citep{SGT_AM} and the He-star channel \citep{SGT_He}. The Evolved CV formation channel is usually given less importance, first because virtually no observed \textcolor{black}{AM CVn stars} have traces of hydrogen in their spectra and it is claimed that the Evolved CV channel would \textcolor{black}{always} leave traces of hydrogen in the system (see \citealt{Nelemans2010} and the references therein). Secondly, the relative importance of this formation channel has been questioned by \cite{2004MNRAS.349..181N} who find that extensive fine-tuning of initial conditions and long time-scales, which exceed the Galactic age, are required to remove all hydrogen from the system. There exist two issues, the lack of H-exhaustion in the system and the fine-tuning problem. However, the main uncertainty with these conclusions about the relative importance of the Evolved CV channel is its strong dependence on the assumed mechanism for $\mathrm{AML_{MB}}$. Previous studies have heavily relied on the empirical magnetic braking formula of \cite{1981A&A...100L...7V} and \cite{1983ApJ...275..713R}. \cite{SGT_AM} models Evolved CV AM CVn stars and finds that the time-scale for $\mathrm{AML_{MB}}$ in our DD model is shorter than that of previously used empirical
formulae. Owing to the shorter time-scales, binaries from a larger parameter space of initial conditions evolve to form AM CVn stars with the DD model than with other models within the Galactic age (see figs~2, 3 and 8 of \citealt{SGT_AM}). This resolves the fine-tuning problem. In Fig.~\ref{fig:PM_SGT} we show how more evolved CVs with bigger H-exhausted cores can be created by starting with a larger initial $P_\mathrm{orb}$. Owing to their larger initial separations, these systems naturally take longer to evolve to typical AM CVn periods $10\lesssim P_\mathrm{orb}/\,\mathrm{min}\lesssim65$ and donor masses $M_2\lesssim0.1M_\odot$. With our strong DD $\mathrm{AML_{MB}}$ these systems can become extremely H-exhausted AM CVn stars within the Galactic age. Owing to their larger H-exhausted cores, their spectra show lower H abundances $X$ at a given $M_2$, to the point of being completely H-exhausted ($X\leq10^{-4}$, see \citealt{Green2019} for an analysis of this constraint) and becoming indistinguishable from the He-star and the WD channel in terms of absence of H in their spectra. This resolves the H-exhaustion problem. The trajectories of AM CVn stars with varying extents of H-exhaustion are plotted in the $P_\mathrm{orb}-M_2$ and $X-M_2$ planes in Figs~\ref{fig:pm-solvan} and \ref{fig:x}. We conclude that well-known AM CVn systems such as YZ~LMi, V396~Hya, CR~Boo and HP~Lib can be explained with the Evolved CV formation channel.

\begin{figure*}
\centering
\includegraphics[width=1\textwidth]{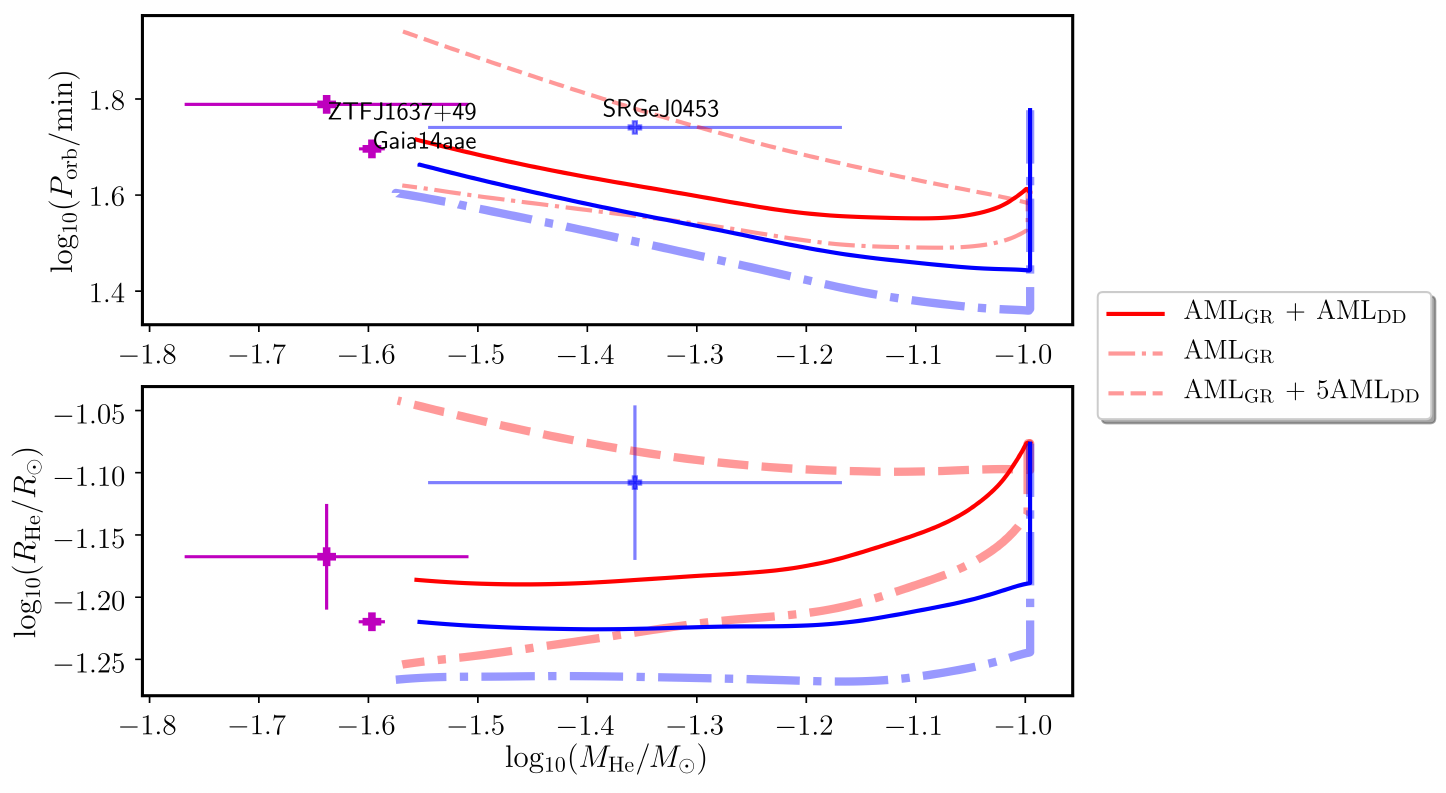}
\caption{Evolutionary tracks of He-stars in the $(\mathrm{log}\,M_\mathrm{He},\,\mathrm{log}\,P_\mathrm{orb})$ and $(\mathrm{log}\,M_\mathrm{He},\,\mathrm{log}\,R_\mathrm{He})$ planes, with a $0.9M_\odot$ WD primary. Starting with the same initial conditions, the solid-line trajectories are evolved with $\mathrm{AML_{GR}}\,+\,\mathrm{AML_{DD}}$, the dashed-line trajectories are evolved with $\mathrm{AML_{GR}}\,+\,5\mathrm{AML_{DD}}$ and the dash-dotted-line trajectories are evolved with only $\mathrm{AML_{GR}}$. ZTFJ1637+49 \citep{2022MNRAS.512.5440V}, Gaia14aae \citep{Green2018} and SRGeJ045359.9+622444 \citep{2023ApJ...954...63R} are observed AM CVn stars with bloated donors.}
\label{fig:PRM_adhoc}
\end{figure*}

We also study AM CVn stars formed through the He-star channel \footnote{It has been pointed out by \cite{2023A&A...678A..34B} that the He-star channel is traditionally defined by a He-burning donor star. This differs from our definition of a He-dominant ($Y\approx0.98$) donor star.} by modelling their evolution after the second CE event (Section~\ref{sec:2step_he_star} in \citealt{SGT_He}). The evolution of these He-star plus WD binaries is driven by $\mathrm{AML_{MB}}$ and $\mathrm{AML_{GR}}$. We argue that the physics governing $\mathrm{AML_{MB}}$ for canonical CVs should still be at play for He-star donors. This is because He-star donors develop a convective envelope as a response to mass loss so now the donor has a convective envelope and a radiative core, the two ingredients required for our DD model to operate. How AML due to MB affects our model trajectories is profound, as shown in Fig.~\ref{fig:PRM_adhoc}, where the dash-dotted lines evolved with just $\mathrm{AML_{GR}}$ are the counterparts to the solid lines evolved with $\mathrm{AML_{GR}}\,+\,\mathrm{AML_{DD}}$. We see that no tracks modelled with $\mathrm{AML_{GR}}$ evolve to match either ZTFJ1637+49 \citep{2022MNRAS.512.5440V}, Gaia14aae \citep{Green2018} or SRGeJ045359.9+622444 \citep{2023ApJ...954...63R}. These are AM CVn systems that possess bloated donors, and as a consequence, have larger $P_\mathrm{orb}$ for their donor mass. In the systems evolved with $\mathrm{AML_{GR}}$ neither the orbital period nor the radius of the donor increase enough during the expansion phase of the system. On the other hand, the systems evolved with $\mathrm{AML_{GR}}\,+\,\mathrm{AML_{DD}}$ clearly show an increasing $P_\mathrm{orb}$ with reducing donor mass which can explain ZTFJ1637+49 and Gaia14aae. This is because the strong mass loss as a result of strong AML by MB gradually makes the donor more bloated for its mass. The time-scales over which $\mathrm{AML_{DD}}$ and $\mathrm{AML_{GR}}$ operate in He-star donors are illustrated in fig.~11 of \cite{SGT_He}. These results suggest that some sort of additional AML mechanism should be incorporated into the modelling of AM CVn stars at such short orbital periods. 

\begin{figure*}
\centering
\includegraphics[width=0.6\textwidth]{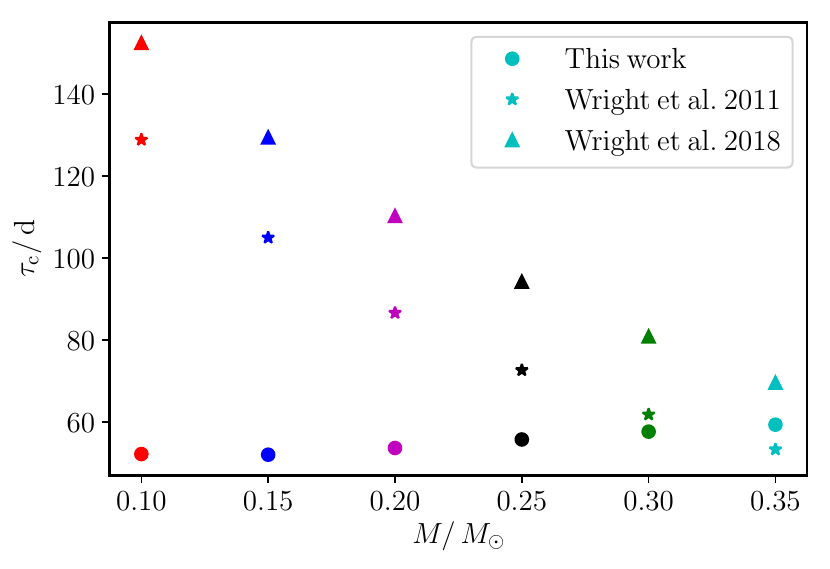}
\caption{The comparison of the estimates of $\tau_\mathrm{c}$ from \cite{SYT} with the estimates from  \citet{2011ApJ...743...48W} and \citet{2018MNRAS.479.2351W} for isolated convective M-dwarfs. Because of the time-dependence of our $\tau_\mathrm{c}$ estimates, we plot $\tau_\mathrm{c}$ from our work for stars beyond their contraction phase when our $\tau_\mathrm{c}$ attains an approximately constant value. This figure is taken from \cite{SYT}. }
\label{fig:SYT_tauc}
\end{figure*}

\begin{figure*}
\centering
\includegraphics[width=0.7\textwidth]{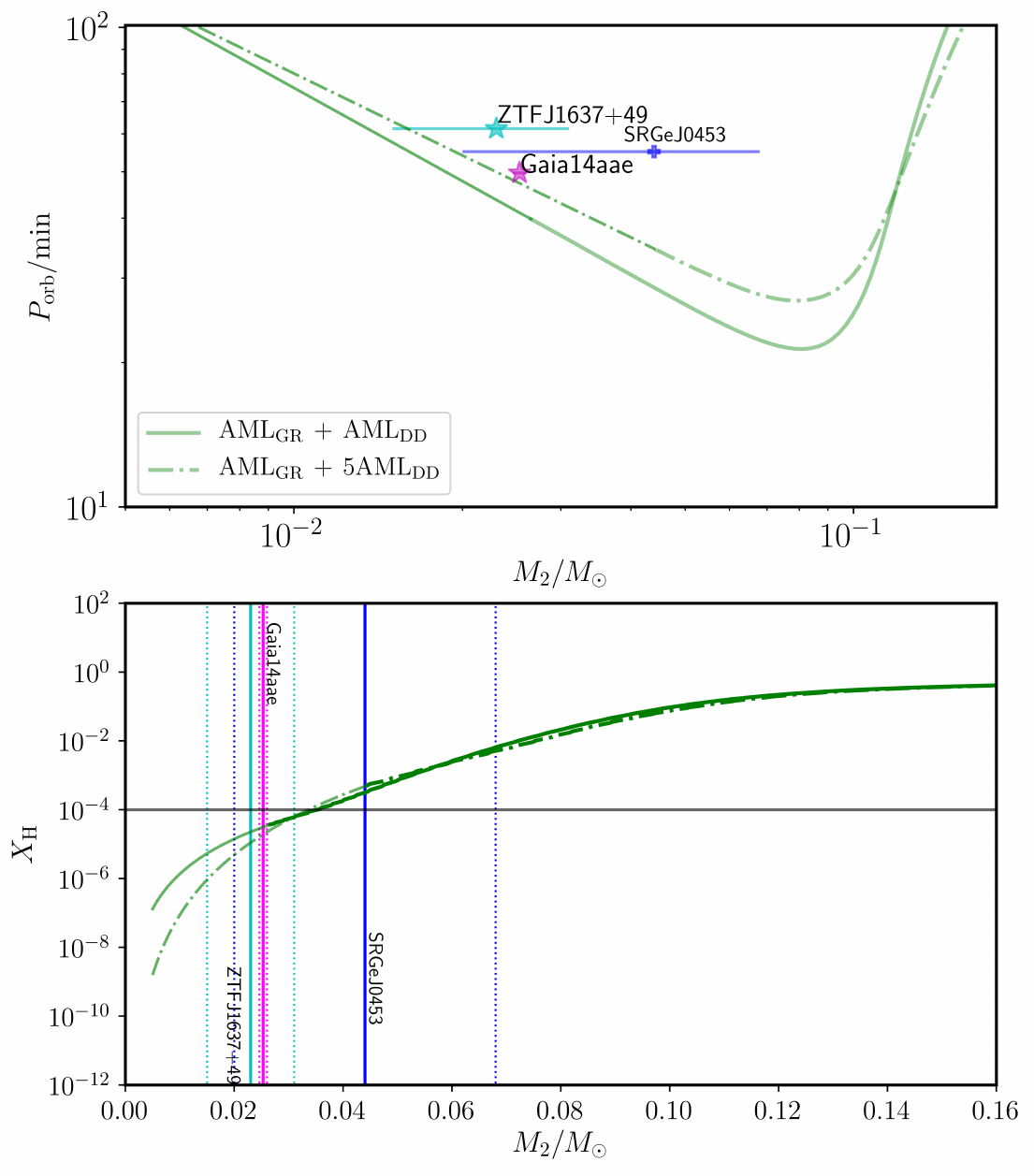}
\caption{Model 3 in green from Figs~\ref{fig:pm-solvan} and \ref{fig:x} with the solid-line trajectory evolved with $\mathrm{AML_{GR}\,+\,AML_{DD}}$ and the dash-dotted line evolved with $\mathrm{AML_{GR}\,+\,5AML_{DD}}$.  ZTFJ1637+49 \citep{2022MNRAS.512.5440V}, Gaia14aae \citep{Green2018} and SRGeJ045359.9+622444 \citep{2023ApJ...954...63R} are observed AM CVn stars with bloated donors.}
\label{fig:PM_palermo}
\end{figure*}

Although uncertainties in modelling the second CE event plague the conclusions of the He-star and the WD channels, any efforts to distinguish between different AM CVn formation channels \citep{Nelemans2010} and their relative importance \citep{2023A&A...678A..34B} most importantly rely on the assumed magnetic braking strength. The strength of MB is heavily model-dependent \citep{1983ApJ...275..713R, Knigge2011, Van2018}, and depends on uncertainties in the assumed value of a parameter within a given model. Let us illustrate this with an example. Various MB formalisms for single- and binary-star evolution, including our DD model, rely on the convective turnover time-scale $\tau_\mathrm{c}$. The semi-empirical formulae of \cite{Matt2015} and \cite{Garraffo2018} for the spin-down of low-mass stars have MB torque $\Dot{J}\propto \tau_\mathrm{c}^2 $ and $\Dot{J}\propto \tau_\mathrm{c} $ respectively, while that of \cite{Van2018} for the evolution of CVs and X-ray binaries has  $\Dot{J}\propto \tau_\mathrm{c}^{8/3}$. Now, $\tau_\mathrm{c}$ for a main-sequence star can be obtained either with the mixing-length theory (MLT) with parameters obtained from a stellar evolution code \citep{SYT, 2023A&A...678A..34B} or with observationally inferred estimates \citep{2011ApJ...743...48W, 2018MNRAS.479.2351W}. These estimates are plotted for single, fully convective M-dwarfs in Fig.~\ref{fig:SYT_tauc}. The $\tau_\mathrm{c}$ estimates from MLT differ from those observationally inferred, with the difference becoming increasingly severe for lower $M$. An additional source of uncertainty in $\tau_\mathrm{c}$ for CV donors is that the donor does not behave like its isolated counterpart and is more luminous depending on the mass-loss rate \citep{Knigge2011}. The $\tau_\mathrm{c}$ for a $M=0.1M_\odot$ star obtained by \cite{2018MNRAS.479.2351W} is about three times that obtained by \cite{SYT}. As CVs, and more importantly, AM CVn stars are observed with $M_2\lesssim0.1M_\odot$, such differences in $\tau_\mathrm{c}$ can strongly increase or decrease the MB strength predicted by a model with a strong $\tau_\mathrm{c}$ dependence. In other words, such MB uncertainties can creep in from several poorly constrained parameters which can lead to noticeable differences in the AM CVn trajectories. This is illustrated in Fig.~\ref{fig:PRM_adhoc} for the He-star channel and Fig.~\ref{fig:PM_palermo} for the Evolved CV channel, where we plot our Evolved CV and He-star AM CVn trajectories respectively assuming that the strength of MB due to the DD model $\mathrm{AML_\mathrm{DD}}$ has been increased by a factor of 5. Fig.~\ref{fig:PRM_adhoc} shows that the $\mathrm{AML_{GR}\,+\,AML_{DD}}$ track fails to reproduce the observed estimates of SRGeJ045359.9+622444 but with $\mathrm{AML_{GR}\,+\,5AML_{DD}}$ we obtain a model trajectory that explains it with the He-star formation channel starting with the same initial conditions. Similarly, Fig.~\ref{fig:x} shows the Evolved CV AM CVn trajectory of Model 3 in Fig.~\ref{fig:pm-solvan}. \cite{SGT_AM} found that our evolved CV AM CVn tracks were not sufficiently H-exhausted to explain either Gaia14aae or ZTFJ1637+49. In Fig.~\ref{fig:x} we show that just with a modest increase in $\mathrm{AML}_\mathrm{MB}$, Model 3 trajectory matches with Gaia14aae, ZTFJ1637+49 and SRGeJ045359.9+622444 (lower limit) in the $P_\mathrm{orb}-M_2$ plane while also being H-exhausted in the $X-M_2$ plane, thereby by explaining their formation through the Evolved CV channel. Thus, we illustrate that claims of whether or not a particular formation channel can justly explain an AM CVn observation strongly depend on the MB mechanism. Detailed orbital parameters and inferred abundance estimates of AM CVn systems will help us better constrain the behaviour of MB at different donor masses and thereby make better predictions of the relative importance of different formation channels.

\section{Towards a holistic magnetic braking model from the evolution of CVs to stellar spin-down}
\label{sec:M-dwarfs}

\begin{figure*}
\centering
\includegraphics[width=1\textwidth]{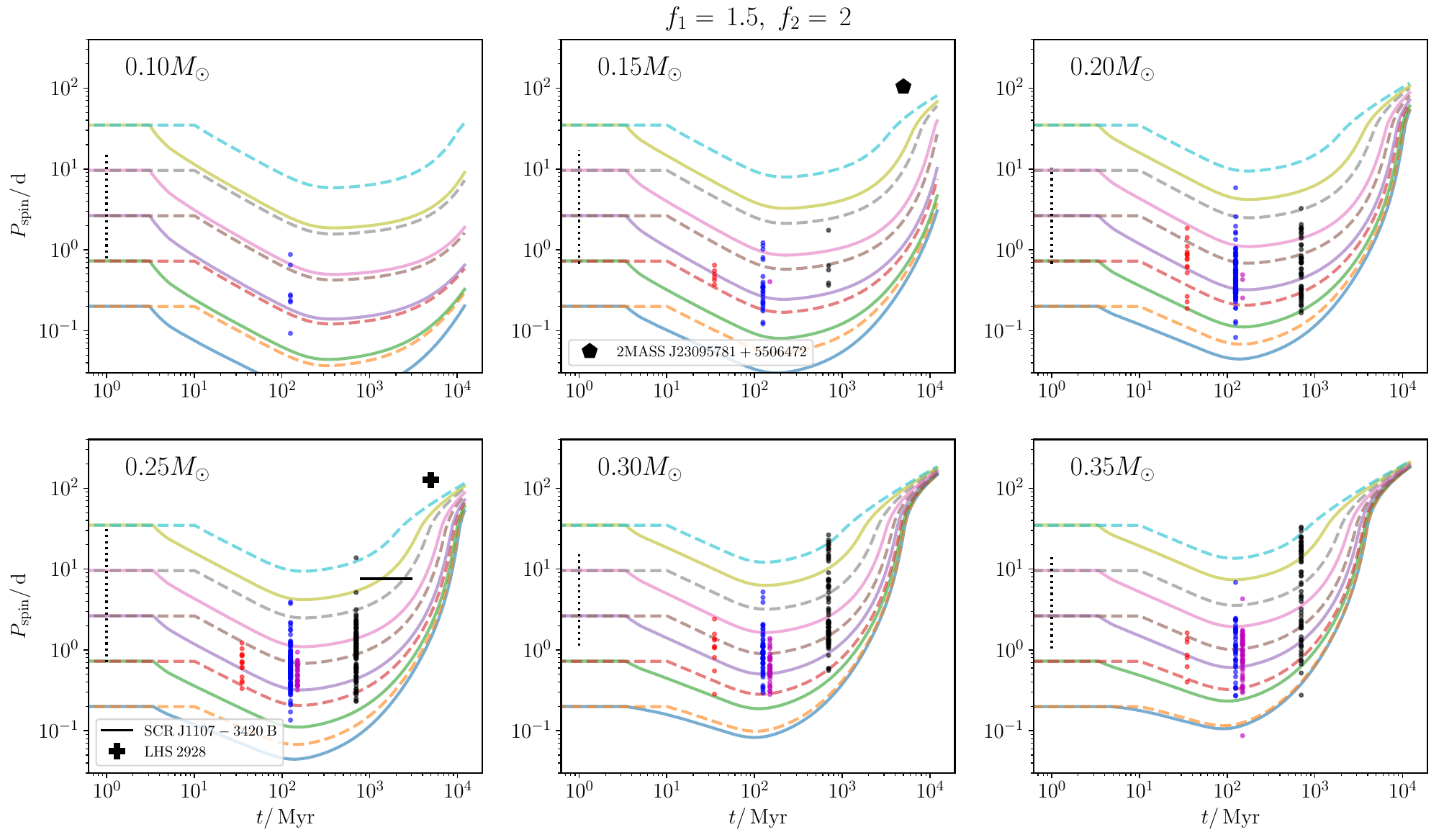}
\caption{Spin evolution 
{of stars} 
with $M_\ast/\,M_\odot \in \{0.1,\,0.15,\,0.2,\,0.25,\,0.3,\,0.35\}$, initial $P_\mathrm{spin}$ equally spaced in log in the range $[0.2,\,35]\,\mathrm{d}$, and $\tau_\mathrm{dl}/\,\mathrm{Myr} \in\{3,\,10\}$ (solid and dotted lines respectively) for $f_1=1.5$ and $f_2=2$ (free parameters discussed in detail in \citealt{SYT}). The observations of OCs   {stars} from \protect\cite{2021ApJS..257...46G} are plotted as a vertical array of dots, such that the dots in red are systems from NGC2547 (about 35 Myr old), dots in blue are Pleiades (about 125 Myr old), dots in magenta are NGC2516 (about 150 Myr old), and dots in black are Praesepe (about 700 Myr old). The dotted line at 1 Myr spans the maximum and minimum spins observed in each mass bin in the Orion Nebula Cluster by \protect\cite{1997AJ....113.1733H}. $\mathrm{2MASS\,J23095781\,+\,5506472}$ and $\mathrm{LHS \,2928}$ with their minimum inferred ages, $\mathrm{SCR\, J1107\, -\, 3420\,B}$ with its age range and spin periods \protect\citep{Pass2022} are also plotted. This figure is taken from \cite{SYT}. }
\label{fig:SYT_PT}
\end{figure*}

Magnetic braking is a process in which magnetized winds from the star carry away mass and, as a consequence, angular momentum. This mechanism also operates in single stars, where it spins them down over time \citep{1967ApJ...148..217W}. Stellar spin-down is heavily dependent on the stellar mass $M_\ast$, such that stars with radiative envelopes ($M_\ast\gtrsim 1.4M_\odot$) do not spin down appreciably over time whereas less massive stars, such as our Sun, with convective envelopes, spin down with time \citep[see, e.g.,][and references therein.]{1992ApJ...390..550M}. Efforts to study MB led to the formulation of the MB torque scaling with spin period as $P_\mathrm{spin}^{-3}$ by \cite{1981A&A...100L...7V} which is widely used in the computation of the orbital evolution of binary stars such as CVs and X-ray binaries. However, modelling the spin-down of single, low-mass stars comes with its own set of challenges. It was soon evident that the MB torque has a weaker dependence on the rotational velocity $\Omega$ in fast-spinning stars \citep{1988ApJ...333..236K, 2011ApJ...743...48W}. A saturated MB torque scaling as $P_\mathrm{spin}^{-1}$ for fast rotators has been incorporated in various MB modelling studies \citep{Sills2000, Matt2015}. Observations of isolated low-mass stars in open clusters (OCs) of known ages also revealed that they show a bimodality in their rotation rates, such that the same OC contains a population of fast as well as slow rotators \citep{2003ApJ...586..464B}. For low-mass stars with radiative cores ($0.3M_\odot\lesssim M_\ast\lesssim 1.4M_\odot$), modelling the surface spins is intricately related to the internal rotational evolution. The problem is two-fold: not only do we require the surface rotation rates of our models to agree with robust observations \citep{2021ApJS..257...46G, Pass2022}, we also require that the core and the envelope corotate by Gyr time-scales in stars with $M_\ast\gtrsim 0.8M_\odot$ \citep{Btrisey2023}. 

\cite{SYT} updated our MB prescription in our DD model to explain various aspects of the spin-down of fully convective M-dwarfs (FCMDs, $M_\ast\lesssim0.35\,M_\odot$). Being fully convective, we only need to carefully model the convective $\alpha-\Omega$ dynamo. The spin-down of our modelled FCMDs is illustrated in Fig.~\ref{fig:SYT_PT}. Modifications in the MB formalism lead to a spin-down torque that scales as $P_\mathrm{spin}^{-1}$ for fast rotators before transitioning to $P_\mathrm{spin}^{-3}$ as the stars spin down. Our MB model can explain the spread in rotation rates in young OCs (ages less than a Gyr) as well as give reasonable estimates for the spins of field stars (ages greater than about a Gyr). For an extensive list of updates implemented in our work we urge the reader to refer to \cite{SYT}. In an ongoing study, we also model the surface and internal spin-evolution of solar-like stars \citep{S_et_al_prep}. In summary, we show that our MB formalism can not only model the orbital evolution of binaries such as CVs and AM CVn stars, but can also model the spin-down of low-mass single stars. This is a crucial step forward towards a MB mechanism that is more robust to uncertainties and that can explain a diverse range of astrophysical systems, such as X-ray binaries \citep{Van2018}, Black widows \citep{2023MNRAS.525.2708C}, etc, in which it is at play. 


\section{Conclusions}
\label{sec:conclusion}

Here we illustrate the results and new insights we have obtained on the orbital evolution of cataclysmic variables (CVs, \citealt{ST}) and AM Canum Venaticorum (AM CVn) systems \citep{SGT_He, SGT_AM} and the spin evolution of single, low-mass stars \citep{SYT} using a mechanism of angular momentum loss (AML) via magnetic braking (MB). Our MB model for binaries is called a double dynamo (DD) mechanism in which MB is driven by two $\alpha-\Omega$ dynamos, one in the convective envelope and the other at the boundary of the convective envelope and the radiative core of the donor star. Our results are summarized below.

\begin{enumerate}
    \item For non-magnetic unevolved CVs, our MB model reproduces the period gap ($2\lesssim P_\mathrm{orb}/\,\mathrm{hr}\lesssim~3$) and the period minimum spike  ($P_\mathrm{orb}\approx 80\, \mathrm{min}$) in their observed distribution. The period gap begins when the donor star abruptly loses its boundary layer, thereby ceasing the strong AML driven by the boundary-layer dynamo. This makes the donor shrink back into its Roche lobe and cease mass transfer. The gap ends when mass transfer resumes following orbital shrinkage driven by AML by the convective dynamo and gravitational radiation. The period minimum happens at $P_\mathrm{orb}\approx 80\, \mathrm{min}$ because of the strong AML by the convective dynamo. It would have occurred at $P_\mathrm{orb}\approx 65\, \mathrm{min}$, in disagreement with observations, if AML were solely due to gravitational radiation.

    \item We find that observed CVs with $P_\mathrm{orb}\geq5.5\,$hr do not match well with our models. In addition, observationally inferred donor mass-radius relations predict larger donor radii than those predicted by stellar evolution models of unevolved CVs. We resolve both these issues by arguing that systems at long orbital periods are CVs with nuclear-evolved donors. These systems are more bloated than their unevolved counterparts and, as a consequence, have larger $P_\mathrm{orb}$ for a given donor mass. They are more luminous and spend more time at longer $P_\mathrm{orb}$. This makes them more likely to be detected at long periods than unevolved CVs. Owing to their similar evolution tracks, it is very difficult to distinguish between evolved and unevolved CVs. The contamination by evolved CVs is the likely cause of the bloated donor mass-radius estimates of long-period CVs. These systems do not detach at the standard CV period gap and so also populate $2\lesssim P_\mathrm{orb}/\,\mathrm{hr}\lesssim~3$.

    \item We study the second common envelope (CE) evolution phase that forms the He-rich donor of AM CVn stars formed through the He-star channel. We find that to form viable post-CE systems, the progenitor star must enter CE before He-ignition at its subgiant or early red giant phase. Owing to the presence of a sizeable radiative region between the convective envelope and the degenerate core at this stage, classical CE formalism is not a viable tool to predict post-CE configurations of such stars. Instead, we consider CE as a two-step process, starting as a canonical CE event that proceeds in a dynamical time-scale and expels the convective envelope, followed by a dynamically stable mass-ejection event, as suggested by \cite{Hirai2022} for massive stars. 

    \item We track the evolution of evolved CVs and post-CE He-star plus white dwarf configurations at very short orbital periods ($10\lesssim P_\mathrm{orb}/\,\mathrm{min}\lesssim~65$) where they become AM CVn stars with the Evolved CV and He-star formation channels respectively. We find that their orbital evolution is driven by AML by GR as well as MB. MB operates by the DD model because these donors still possess a convective envelope and a radiative core. For the Evolved CV channel we find that, owing to strong $\mathrm{AML_{MB}}$, binaries from a larger parameter space of initial configurations evolves to form AM CVn stars within the Galactic age. This solves the fine-tuning problem of this channel. Even the most evolved-donor systems, that take the longest time to evolve, form AM CVn stars within the Galactic age. They become extremely H-exhausted systems. This makes them indistinguishable from systems evolved from the He-star and the White Dwarf (WD) channels in terms of the absence of H in their spectra. This solves the H-exhaustion problem. Overall, we show that the Evolved CV channel is important and several well-known AM CVn stars could have formed through this channel. Similarly, for the He-star channel the donors experience strong mass loss driven by MB in our DD model which causes them to bloat up and, as a result, expand their orbit. This can explain the existence of AM CVn stars with bloated donors such as Gaia14aae and ZTFJ+1637+49. These systems cannot be explained with models evolved solely with GR. Overall, this suggests the importance of incorporating some sort of MB physics in the evolution of AM CVn stars at such short periods.

    \item We describe uncertainties in various MB formalisms and how these can lead to profound effects in the efforts to distinguish between different AM CVn channels and their relative importance. We illustrate this by increasing our MB strength by an ad hoc factor, which makes all our AM CVn donors more bloated, thereby explaining Gaia14aae, ZTFJ+1637+49 and SRGeJ045359.9+622444 with both the formation channels. 

    \item Since MB also drives stellar spin-down, we make our MB formalism, in particular the convective dynamo, more robust so that it also explains the spin-down of single, fully convective M-dwarf stars. We discuss our plans to model the spin-down of all low-mass stars with convective envelopes in the future.
\end{enumerate}

\begin{acknowledgments}
AS thanks the Gates Cambridge Trust for his scholarship. HG acknowledges the National Natural Science Foundation of China (NSFC, grant Nos. 12288102, 12090040/3, 12173081), National Key R\&D Program of China (2021YFA1600403), Yunnan Fundamental Research Projects (grant NOs. 202101AV070001), the Key Research Program of Frontier Sciences, CAS, No. ZDBS-LY-7005. CAT thanks Churchill College for his fellowship. 
\end{acknowledgments}

\bibliographystyle{mnras}
\bibliography{skeleton} 
\bigskip
\bigskip
\bigskip
\noindent {\bf DISCUSSION}

\bigskip
\noindent {\bf KEN SHEN:} Do you use the same normalization factors in your MB for both evolved and unevolved CVs?

\bigskip
\noindent {\bf ARNAB SARKAR:} Yes, all the results discussed in our works use the same set of normalization factors for unevolved CVs, evolved CVs and AM CVn stars. Even the free parameters in our spin-down work stay the same for every star.
\\
..............

\bigskip
\noindent {\bf MARIKO KIMURA:} In the two-step CE model, stable mass transfer occurs. What kind of transients correspond to this stage?

\bigskip
\noindent {\bf ARNAB SARKAR:} We have not done any analysis on the nature of transients in our work.

\end{document}